\documentstyle[times,pramana,epsf,floats]{ias}

\def\sst{\scriptscriptstyle}
\def\as{\alpha_{\sst S}}
\def\msbar{\overline {\rm MS}}
\newcommand\mh{m_{\sst H}}
\def\beq{\begin{equation}}
\def\eeq{\end{equation}}
\def\bea{\begin{eqnarray}}
\def\eea{\end{eqnarray}}

\begin{document}
\mark{{Vittorio Del Duca}{QCD at hadron colliders}}
\title{QCD at hadron colliders}

\author{Vittorio Del Duca}
\address{Istituto Nazionale di Fisica Nucleare, Sez. di Torino\\
via P. Giuria, 1 - 10125 Torino, Italy}
\abstract{
QCD is an extensively developed and tested gauge theory, which models
the strong interactions in the high-energy regime. In this talk, I shall
review the considerable progress which has been achieved in the last
few years in the most actively studied QCD topics: Monte Carlo models,
higher-order corrections, and parton distribution functions. 
Thanks to that, QCD in the high-energy regime
is becoming more and more an essential precision toolkit to analyse
Higgs and New Physics scenarios at the LHC.}

\maketitle
\section{Introduction}

Quantum chromodynamics (QCD)
is by now widely accepted as the theory which describes the strong
interactions between hadrons and their components, the quarks and gluons.
From a theoretical point of view,
it is a gauge field theory featuring asymptotic
freedom, i.e. a coupling that grows weaker at smaller distances.
However, the strong interactions also feature confinement, that is
the lack of colour of the observed hadrons. Although
in the non-perturbative (low $Q^2$) regime several clever approaches to QCD,
like lattice gauge theory, Regge theory, chiral perturbation theory,
large $N_c$, are used, a complete theoretical
solution to confinement is not yet available and is difficult to
obtain, because the QCD Lagrangian is formulated in terms of quarks
and gluons, rather than the observed hadrons, and because at large 
distances, {\it i.e.} at low $Q^2$, the coupling is strong.
Conversely, at small distances, i.e. in the high-energy regime, the
coupling is weak and thus it is possible to make use of the
perturbative framework. In this talk
I shall focus on the fact that perturbative QCD (pQCD), 
{\it i.e.} QCD in the weak-coupling (high $Q^2$) regime, has emerged as an 
essential precision toolkit for exploring Higgs and Beyond-the-Standard-Model
(BSM) physics; and that
is even more so at the Large Hadron Collider (LHC), because of the
strongly interacting colliding protons.
In the box of the precision toolkit, pQCD provides the tools for a
precise determination of the strong coupling constant, $\alpha_{\sst S}$,
of the parton distribution functions (p.d.f.), of the electroweak parameters 
and of the LHC parton luminosity; and of the
strong corrections to Higgs and BSM signals and to their backgrounds.

In any scattering process in
high-energy QCD, the value of any observable can be expanded in principle
as a series in $\as$. Thus $\as$ represents the single most important piece
of information we need. In the $\msbar$ scheme and using
next-to-next-to-leading-order (NNLO) results only, the 2004 world 
average~\cite{Bethke:2004uy} yields $\as(M_Z) = 0.1182\pm 0.0027$.

The fact that in the detectors experiments observe hadrons, while through
the QCD Lagrangian we can only compute the scattering between partons, calls
for a framework where the short-distance physics, which is responsible for
the primary scattering between partons, can be separated from the long-distance
physics, which describes the parton densities in the initial state and the
hadronisation in the final state. That framework is provided by the pQCD
and factorisation.

In the beginning there was the parton model and the Bjorken scaling,
which constitute the backbone of pQCD. The latter computes
the logarithmic scaling violations, that is the logarithmic corrections
to the parton-model predictions. The pQCD tenets are the universality of
the infrared (IR) behaviour, the cancellation of the IR singularities
for suitably defined variables, like jets
and event shapes, and in the case of hadron-initiated processes, like
electron-proton or (anti)proton-proton collisions, the factorisation of
the short- and long-range interactions. 

Factorisation in proton-proton ($pp$) collisions states that the
cross section for the production of high-mass states, like vector bosons,
Higgs bosons, heavy quarks or large-$E_T$ jets, characterised by the
large scale $Q^2$, can be expressed as a convolution
of short- and long-distance pieces, with the matching between the two pieces
occurring at an arbitrary scale $\mu_F$, called factorisation scale.
The short-distance piece is the parton cross section
for the primary event, describing how two partons out of the incoming protons
collide to produce the high-mass states plus hard radiation;
the long-distance pieces are the p.d.f.'s in the initial state (and 
eventually the fragmentation functions in the final state),
describing the density of the colliding partons within the
protons, and how that density changes with the parton virtuality and
momentum fraction. The p.d.f.'s cannot be computed -- they must be given by the 
experiment -- but their dependence on $\mu_F$ can.
If factorisation holds, the parton cross section may be computed 
as a power series in $\alpha_{\sst S}$. The only limitation is then 
computational,
and the state of the art is that some production rates with one (and very few
with two) final-state particles are known at next-to-next-to-leading order
(NNLO), while many with two and some with three (and for special cases with
four) final-state particles are
known at next-to-leading order (NLO). A matching accuracy is then required
from the p.d.f.'s whose DGLAP evolution has been recently computed to 
NNLO~\cite{Moch:2004pa,Vogt:2004mw}.

\section{Breaking factorisation}\label{sec:fact}

Factorisation, though, ignores altogether the underlying event (UE),
which can be operatively defined as whatever is in the $pp$ interaction
besides the primary scattering. In particular, the UE
includes the multiple-parton interactions as well as the interaction
of spectator partons, {\it i.e.} other than the ones initiating the primary 
scattering, the assumption
being that if such interactions occur they are characterised by a scale 
$\Lambda$ of the order of a GeV, and so they are suppressed by powers of 
$\Lambda^2/Q^2$ with respect to the primary scattering.
Thus, the UE breaks factorisation by means of power-suppressed
contributions. How important are they for a precision calculation ?
There is no obvious answer to this question since of course we cannot use
the pQCD framework to model the UE, and its analysis must rely solely upon 
the data. In $p\overline{p}$ collisions at the Tevatron, the UE is being 
studied~\cite{Acosta:2004wq} by analysing in single-jet production
the charged-particle multiplicity in regions 
which are perpendicular in azimuth to the jet, since that region
is expected to be sensitive to the UE. The UE sensitivity to
beam remnants and to multiple interactions can be reduced by selecting
back-to-back two-jet topologies. A similar investigation is 
being planned also through the Drell-Yan production of vector bosons.

Other examples of factorisation-breaking contributions are: $a)$ the power
corrections: Monte Carlo (MC) and theory modelling of power corrections were 
laid out and tested at LEP, where they also provided a determination of 
$\alpha_S$~\cite{Dokshitzer:1995qm}. However, 
those models still need be tested in 
hadron collisions: a study of single-jet production at the Tevatron running at
two different centre-of-mass energies shows that the Bjorken scaling is violated
more than logarithmically: data can fitted by assuming a power-correction
shift in the jet $E_T$~\cite{santabarb}; $b)$ diffractive 
events~\cite{Collins:1992cv}, which are known to violate factorisation
at the Tevatron~\cite{Terashi:2000vs,Acosta:2003xi}.
\bigskip
\medskip

\begin{figure}[htbp]
\epsfxsize=6cm
\centerline{\epsfbox{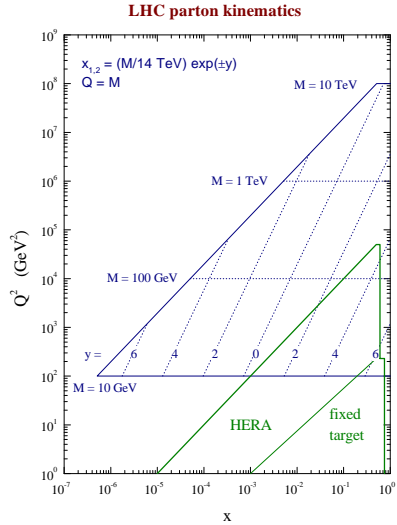}}
\caption{LHC kinematic range, from MRST~[10].}
\label{fig:lhcrange}
\end{figure}

\section{Monte Carlo models}
\label{sec:mc}

The detection of Higgs and BSM signals requires a precise modelling of
their backgrounds. Examples are QCD production of $W + 4$ jets and
of $W W + 2$ jets, which are backgrounds to Higgs production through
vector-boson fusion (VBF) with the Higgs decay into a $W W$ pair, as well
to $t\overline{t}$ production, or $W + 6$ jets and $W W + 4$ jets,
which are backgrounds to $H t\overline{t}$ production. 
The huge amount of phase space available at the 
LHC, Fig.~\ref{fig:lhcrange}, as well as the large acceptance of the detectors,
will make possible to produce final states with ten jets or more.
One approach is to model QCD production through matrix-element MC
generators, which provide an automatic computer generation of
processes with many jets, and/or vector or Higgs bosons.
There are several such multi-purpose generators, like e.g. 
ALPGEN~\cite{Mangano:2001xp,Mangano:2002ea},
ARIADNE~\cite{Lonnblad:1992tz},
MADGRAPH/MADEVENT~\cite{Stelzer:1994ta,Maltoni:2002qb}, 
COMPHEP~\cite{Pukhov:1999gg}, GRACE/GR@PPA~\cite{Ishikawa:1993qr,Yuasa:1999rg}, 
HELAC~\cite{Kanaki:2000ey}, and SHERPA~\cite{Krauss:2004bs} (which has got its 
own showering and hadronisation).
A different example is PHASE/PHANTOM~\cite{Accomando:2004my}, 
a MC generator dedicated to processes
with six final-state partons only -- thus suitable to $t\bar{t}$ production,
$W W$ scattering, Higgs production via VBF and 
vector-boson gauge coupling studies, but where no approximation is used.
Matrix-element MC generators are particularly suitable to studies which
involve the geometry of the event, because the jets in the final state are
generated at the matrix-element level, and thus exactly at any angle.
In addition, they can be interfaced to parton-shower MC generators, 
like HERWIG~\cite{Marchesini:1991ch} or PYTHIA~\cite{Sjostrand:1993yb}, to
include showering and hadronisation. 
In the context of matrix-element generators, 
the presence of several hard scales, like the mass of
high-mass states, the jet transverse energy and the dijet invariant masses,
makes an operative implementation of factorisation rather involved.
The issue can be tackled through the CKKW~\cite{Catani:2001cc,Krauss:2002up}
(or CKKW-like~\cite{Hoche:2006ph}) procedure.
Within a given jet cross section, CKKW interfaces parton
subprocesses with a different number of final states to parton showers.

Finally, MC@NLO~\cite{Frixione:2002ik}: a procedure and a code to match exact 
NLO computations to shower MC generators. In a way, this is the most desirable
procedure, because it embodies the precision of NLO parton calculations
in predicting the overall normalisation of the event,
while generating a realistic event set up through showering and hadronisation.
It cannot be, though, multi-purpose, being obviously limited to the 
processes for which the NLO corrections are known. 

\section{NLO calculations}
\label{sec:nlo} 

Another approach to the evaluation of the parton cross section $\hat\sigma$
is through
fixed-order computations. These yield only a limited access to the
final-state structure, but have the advantage that higher-order
corrections, real and virtual, can be included exactly. The virtual
corrections will depend on a fictitious scale $\mu_R$, at which the
scattering amplitudes are renormalised to take care of the
ultraviolet divergences.
NLO calculations have several desirable features. $a)$ the jet structure:
while in a leading-order
calculation the jets have a trivial structure because each parton becomes
a jet, to NLO the final-state collinear radiation allows up to two partons to
enter a jet; $b)$ a more refined p.d.f. evolution through the initial-state
collinear radiation; $c)$ the opening of new channels, through the inclusion
of parton sub-processes which are not allowed to leading order; $d)$ a
reduced sensitivity to the fictitious input scales $\mu_R$ and $\mu_F$
allows to predict the normalisation of physical observables, which
is usually not accurate to leading order. That is the first step toward
precision measurements in general, and in particular toward an accurate
estimate of signal and background for Higgs and New Physics at LHC;
$e)$ finally, the matching with a parton-shower MC generator, like
MC@NLO, as mentioned in Sect.~\ref{sec:mc} .

Here we remind briefly how a NLO calculation is organised. 
Let us consider the production of $n$ jets
in hadron collisions. There are two types of contributions to $\hat\sigma$:
the tree-level production with $n+1$ final-state partons,
with one of the partons that is undetected,
and the one-loop production with $n$ final-state partons. Schematically,
\beq
\hat\sigma = \sigma^{\rm LO} + \sigma^{\rm NLO}
= \int_n d\sigma^B + \sigma^{\rm NLO}
\eeq
where $d\sigma^B$ is the Born cross section, and
\beq
\sigma^{\rm NLO} = \int_{n+1} d\sigma^R + \int_n d\sigma^V\ .
\label{eq:due}
\eeq
Both real and virtual contributions to Eq.~(\ref{eq:due}), contain IR, 
i.e. collinear and soft, singularities. If in order to regulate those 
divergences one uses the dimensional regularisation, which fixes the dimensions
of space-time to be $d=4-2\epsilon$, then one finds that both terms on
the right-hand side of Eq.~(\ref{eq:due}) are divergent at $d=4$.
However, the structure of QCD is such that those
singularities are {\it universal}, i.e. they do not depend on the
process under consideration, but only on the partons involved in
generating the singularity. Thus, in the 90's process-independent
procedures were devised to regulate those divergences. They are
conventionally called {\it slicing}~\cite{Giele:1991vf,Giele:1993dj}, 
{\it subtraction}~\cite{Frixione:1995ms,Nagy:1996bz}
and {\it dipole subtraction}~\cite{Catani:1996vz}, and use universal 
counterterms to subtract the divergences. The NLO contribution,
Eq.~(\ref{eq:due}), can be written as, 
\beq
\sigma^{\rm NLO} = \int_{n+1} \left[(d\sigma^R)_{\epsilon=0} -
(d\sigma^A)_{\epsilon=0}\right] + 
\int_n \left(d\sigma^V + \int_1 d\sigma^A \right)\ ,
\label{eq:tre}
\eeq
such that both sums of bracketed terms on the right-hand side of 
Eq.~(\ref{eq:tre}) are finite at $d=4$, and thus readily integrable
numerically via a computer code, with arbitrary selection cuts on
the final-state particles and jets, as eventually required by a 
detector simulation. The organisation of NLO computations in
process-independent procedures has made them an essential
tool in the comparison with the experimental data. 

Let us look briefly at the history of NLO calculations:
the first final-state distribution to NLO was computed for
$e^+e^-\to$~3~jets~\cite{Ellis:1980wv}. The addition of just one more jet in the
final state, to produce $e^+e^-\to$~4~jets, took about 15 
years~\cite{Dixon:1997th,Nagy:1997yn}. This trend, namely
the great difficulty in adding one more jet to a given final-state distribution,
is repeated in all
the other NLO calculations: in Drell-Yan with one associated 
jet~\cite{Giele:1993dj}, and with two associated 
jets~\cite{Campbell:2002tg}; in one- or two-jet production in hadron 
collisions~\cite{Giele:1993dj,Ellis:1990ek}, and
in three-jet production~\cite{Kilgore:1999qg,Nagy:2001fj}; 
in di-photon production
in hadron collisions~\cite{Bailey:1992br,Binoth:1999qq}, and in the same 
with one associated jet~\cite{DelDuca:2003uz}. Conversely, 
for other distributions in hadron collisions, 
like heavy-quark pair production~\cite{Mangano:1991jk}, vector-boson pair 
production (including the spin 
correlations)~\cite{Campbell:1999ah,Dixon:1999di}, Drell-Yan with a 
heavy-quark pair~\cite{Campbell:2002tg},
Higgs production via gluon fusion with one associated 
jet~\cite{deFlorian:1999zd,Glosser:2002gm}, Higgs production via vector-boson
fusion with two associated jets~\cite{Figy:2003nv}, Higgs production with a 
heavy-quark pair~\cite{Beenakker:2001rj,Reina:2001sf},
the addition of just one more jet has not been achieved yet to NLO.
Furthermore, all the hadron-initiated distributions above have no more 
than three final-state particles (only very special cases with four 
final-state particles, like the NLO 
corrections to the electroweak production of a vector-boson pair $+$ 2 
jets~\cite{Jager:2006zc}, are known).

Why in a NLO calculation is it so difficult to add more 
particles in the final state ? The loop integrals occurring in the
virtual contributions to Eq.~(\ref{eq:due}) are involved and process dependent.
In addition, more final-state particles imply more scales in the process,
and so lenghtier analytic expressions in the loop integral.
In fact, the only known complete one-loop amplitudes with hexagon loops
are the ones for six 
gluons~\cite{Bern:1994zx,Bidder:2004tx,Bidder:2004vx,Britto:2005ha,Bern:2005hh,Britto:2006sj,Ellis:2006ss,Berger:2006ci}
and for the electroweak corrections to 
$e^+e^-\to 4$~fermions~\cite{Denner:2005es}; with heptagon loops
are the ones for seven gluons in N=4 super-yang-Mills theory~\cite{Bern:2004ky}.
Recently, a twistor-inspired approach~\cite{Witten:2003nn}, which has
allowed for great advances in the analytic computation of tree and one-loop
amplitudes~\cite{Cachazo:2004kj,Britto:2005fq,Bern:2005hs,dak}, 
as well as several semi-numerical approaches which show promise to handle NLO 
corrections in an automated 
way~\cite{Kramer:2002cd,Binoth:2002xh,Nagy:2003qn,Giele:2004iy,Binoth:2005ff,Ellis:2005qe,Anastasiou:2005cb,Czakon:2005rk,Binoth:2006rc}, 
have appeared. However, the programme
of applying sistematically NLO computations to studies of signals and
backgrounds for Higgs and New Physics is still in its infancy
and will be undoubtedly receive a lot of attention in the next future.

\section{NNLO calculations}
\label{sec:nnlo}

Are NLO computations accurate enough to describe the data ? The answer to that
question is of course process dependent. Here I shall give a few examples:
\begin{itemize}
\item {\it $b$ production at the Tevatron} It has been long thought that
the CDF data for $b$ quark production were not in agreement with the NLO
prediction (for a historical overview, see Ref.~\cite{Mangano:2004xr}). 
However, in the comparison of the CDF Run II data~\cite{Acosta:2004yw}
for the $J/\psi$ 
momentum distribution in inclusive $B\to J/\psi + X$ decays to the NLO 
prediction~\cite{Cacciari:2003uh} and to MC@NLO~\cite{Frixione:2003ei}, 
one finds that the data lie
within the theory uncertainty band and are in good agreement with the theory 
predictions.
\item {\it $W$ production at the LHC} The Drell-Yan $W$ cross section,
with leptonic decay of the $W$ boson, has been proposed as a luminosity monitor
of the LHC~\cite{Dittmar:1997md}, warranting a greater accuracy, of the order 
of a few percent, than the standard 
determination of the luminosity through the total hadronic cross section. 
However, the experimental $W$ cross section depends on the acceptance,
i.e. the fraction of events which pass the selection cuts. Thus,
the accuracy of the luminosity monitor, the {\em standard candle}, 
depends on the one of the acceptance,
which is related to the precision by which the hard cross section is known.
In Ref.~\cite{Frixione:2004us} the $W$ cross section has been computed to
different accuracies: to leading order, the same $+$ HERWIG, NLO, MC@NLO,
with or without 
including the spin correlations between the decay leptons and the
partons entering the hard scattering. 
It was found that the difference between the
NLO calculation and MC@NLO is about $2-3\%$, which is much less than
the difference between the same calculations with and without spin
correlations. Therefore, to whatever accuracy we may compute the
$W$ cross section, if we want to use it as a standard candle it is mandatory
to include the spin correlations. Recently, a calculation of the NNLO
corrections, including the spin correlations, has been 
completed~\cite{Melnikov:2006di}. It shows that the NNLO corrections
differ from the NLO corrections more than $2-3\%$ only if severe 
acceptance cuts on the $p_T$ of the outgoing electron are used, which
restrict drastically the available phase space.
\item {\it Higgs production at the LHC} At hadron colliders, the leading
production mode for the Higgs is via gluon-gluon fusion through the 
mediation of a heavy-quark (mostly top-quark) loop.
The NLO corrections to fully inclusive Higgs production via gluon-gluon 
fusion, including the heavy-quark mass dependence, required an evaluation 
at two-loop accuracy, and were found to be
as large~\cite{Graudenz:1992pv,Spira:1995rr} as the leading-order 
calculation. That situation was unsatisfactory, because it
called for a calculation to NNLO~\footnote{One must keep in mind that the
calculation of Ref.~\cite{Graudenz:1992pv,Spira:1995rr} is fully inclusive,
thus for an ideal detector with a $4\pi$ coverage. If selection
cuts are applied, like in Ref.~\cite{Anastasiou:2004xq}, where Higgs 
production via gluon-gluon 
fusion is computed to NLO and to NNLO with a jet veto, the higher-order
corrections may be not so large as in the fully inclusive calculation.
Thus the ultimate judgement on the usefulness of a NNLO evaluation rests 
on an analysis with the cuts which will be 
used in the realistic simulations of the ATLAS and CMS detectors.},
which requires an 
evaluation at three-loop accuracy. However, in the large-$m_t$ limit,
{\it i.e.} when the Higgs mass is smaller than the threshold for the 
creation of a 
top-quark pair, $\mh \lesssim 2 m_t$, the coupling of the Higgs to the
gluons via a top-quark loop can be replaced by an effective 
coupling. That reduces the number of loops in a given diagram by one. 
The NNLO corrections have been evaluated in the 
large-$m_t$ limit~\cite{Harlander:2002wh,Anastasiou:2002yz,Ravindran:2003um}
and display a modest increase with respect to the NLO evaluation, showing 
that the calculation stabilises to NNLO.
\end{itemize}

In the examples above I stressed how the central value of a prediction
may change when going from leading order to NLO and eventually to NNLO.
However, a benefit of going from leading order to NLO and then to NNLO is the
reduction of the theory uncertainty band, due to the lesser
sensitivity to the fictitious input scales $\mu_R$ and $\mu_F$ of the
calculation. Thus, a lot of theoretical activity has been directed in the
last years toward the calculation of cross sections to NNLO accuracy.
The total cross section~\cite{Harlander:2002wh,Hamberg:1990np} and the
rapidity distribution~\cite{Anastasiou:2003yy,Anastasiou:2003ds} 
for Drell-Yan $W, Z$ 
production are known to NNLO accuracy. So are the total cross 
section~\cite{Harlander:2002wh,Anastasiou:2002yz,Ravindran:2003um} 
and the rapidity 
distribution~\cite{Anastasiou:2004xq} for Higgs production via gluon-gluon
fusion, in the large-$m_t$ limit. However, only the calculation of
Ref.~\cite{Anastasiou:2004xq}, which has been extended to include the
di-photon background~\cite{Anastasiou:2005qj}, and Ref.~\cite{Melnikov:2006di}
allow the use of arbitrary selection cuts.

Basically, there are three ways of computing NNLO corrections: 
\begin{itemize}
\item Analytic integration, which is the first method to have been 
used~\cite{Hamberg:1990np}, and may include a limited
class of acceptance cuts by modelling cuts as 
``propagators''~\cite{Anastasiou:2003yy,Anastasiou:2002wq}.
Besides total cross sections, it has 
been used to produce the NNLO differential rates of 
Ref.~\cite{Anastasiou:2003ds}.
\item Sector decomposition, which is flexible enough to include any
acceptance 
cuts~\cite{Roth:1996pd,Binoth:2000ps,Heinrich:2002rc,Anastasiou:2003gr}, 
and has been used to produce the NNLO differential rates of 
Refs.~\cite{Melnikov:2006di,Anastasiou:2004xq,Anastasiou:2005qj}
and of $e^+e^-\to 2$~jets~\cite{Anastasiou:2004qd}.
The cancellation of the IR divergences is performed numerically.
\item Subtraction, for which the cancellation of the divergences is  
organised in a process-independent way by exploiting the universal  
structure of the IR divergences, in particular  
the universal structure of the three-parton tree-level splitting  
functions~\cite{Berends:1988zn,Gehrmann-DeRidder:1997gf,Campbell:1997hg,Catani:1998nv,DelDuca:1999ha}  
and of the two-parton one-loop splitting   
functions~\cite{Bern:1994zx,Bern:1998sc,Kosower:1999xi,Kosower:1999rx,Bern:1999ry}. 
Although, the universal splitting functions have been known for some time,
the cancellation of the IR divergences to NNLO is very 
intricate~\cite{Kosower:2002su,Weinzierl:2003fx,Gehrmann-DeRidder:2003bm,Gehrmann-DeRidder:2004tv,Gehrmann-DeRidder:2005hi,Frixione:2004is,Somogyi:2005xz,Gehrmann-DeRidder:2005cm,Weinzierl:2006ij}, and except for test cases
like $e^+e^-\to 2$~jets~\cite{Gehrmann-DeRidder:2004tv,Weinzierl:2006ij} and 
for parts of $e^+e^-\to 3$~jets~\cite{Gehrmann-DeRidder:2005cm}, 
no NNLO numerical code has been devised yet.
\end{itemize}

\section{The parton distribution functions}
\label{sec:pdf}

As outlined in the Introduction, at hadron colliders the theory cross
section can be written using factorisation as a convolution of the 
parton cross section
with the p.d.f.'s. The dependence of the p.d.f.'s on
$Q^2$ is given by the DGLAP evolution equations. In those equations,
the evolution in $Q^2$ is driven by the splitting functions,
which are perturbatively computable. By consistency, in the factorisation 
formula the parton cross section and the splitting functions must be
determined to the same accuracy. The 
leading-order~\cite{Gross:1973ju,Altarelli:1977zs} and
NLO~\cite{Curci:1980uw} splitting functions have been known for a long time.
The calculation of the NNLO splitting functions has been completed 
recently~\cite{Moch:2004pa,Vogt:2004mw}, setting the record as
the toughest calculation ever performed in perturbative QCD: it took the
equivalent of 20 man-years, and about a million lines of dedicated 
algebra code.
The p.d.f.'s obtained by global fits~\cite{Martin:2002dr}\footnote{In
Ref.~\cite{Martin:2002dr}, which pre-dates Refs.~\cite{Moch:2004pa,Vogt:2004mw},
the NNLO global fit is based on a few NNLO fixed moments, which were
known at that time.} of all 
accessible collider and fixed-target data can be evolved to the large 
kinematic range accessible through the LHC.
In global fits, the fit is performed by minimising the $\chi^2$ to
all the data. The evolution is started at some value $Q^2_0$, where
the p.d.f. is some suitable function of $x$. In addition, to avoid higher-twist
contaminations, the data are selected above a certain momentum transfer
and energy, $Q^2 > Q^2_{\rm min}$ and $W^2 > W^2_{\rm min}$.
Recently, though, also an evaluation of the $\Delta\chi^2$, i.e. of the 
uncertainties arising from the errors on the experimental data, has been 
performed~\cite{Stump:2001gu,Pumplin:2001ct,Pumplin:2002vw,Martin:2002aw,Martin:2003sk}, 
using either the Hessian or the Lagrange-multiplier 
methods\footnote{
Accordingly, in connection with the use of $W, Z$ production as
a parton luminosity monitor mentioned in Sect.~\ref{sec:nnlo},
Ref.~\cite{Martin:2003sk} estimates a $4\%$ uncertainty on the
Drell-Yan $W, Z$ production cross section.}.

\section{Conclusions}

QCD in the high-energy regime is constantly making progress.
Here I have reviewed the considerable advances achieved in the last
few years in the sectors of QCD which are
most actively studied: Monte Carlo models, higher-order corrections
and p.d.f.'s. More progress can be anticipated in modelling the
underlying event, in improving the factorisation picture in Monte Carlo
models through the use of CKKW-like procedures, in the automatisation
of NLO calculations, and in a working NNLO numerical code based on
subtraction, thus making high-energy QCD 
more and more an essential precision toolkit to analyse
Higgs and New Physics scenarios at the LHC.

\section*{Acknowledgments}

I should like to thank James Stirling for providing the figure used in the 
text, and the organisers of WHEPP9 for their kind hospitality.

\end{document}